# ARE CLUSTERS AS INDICATORS OF THE COSMIC RAY ANISOTROPY ?

A.A.Mikhailov, N.N.Efremov

Yu.G. Shafer Institute of Cosmophysical Research and Aeronomy, 31 Lenin ave., 677890 Yakutsk, Russia

**Abstract:** The clusters (doublets) in ultrahigh energy cosmic rays are considered based on Yakutsk and AGASA extensive air shower array data. The problem of cluster origin is discussed. It is found that arrival directions of the clusters can point to a cosmic ray anisotropy. As a result of analysis of clusters, the conclusion on composition and origin of ultrahigh energy cosmic rays has been made.

## 1. INTRODUCTION

In the ultrahigh energy region at $E>4.10^{19}$ eV 7 clusters were found by AGASA extensive air shower (EAS) array data[1], 2 clusters by Yakutsk EAS array data[2]. A mechanism for the appearance of clusters remains unsolved. It is supposed[1] that showers at $E>4.10^{19}$ eV corresponding to the clusters are formed by the neutral particles arriving from cosmic ray sources. We assumed[2] that the clusters are formed as a result of fragmentation of super heavy nuclei into more light ones.

## 2. EXPERIMENTAL DATA

According to the hypothesis[2], among the clusters, forming as result of fragmentation, the most energetic particle must arrive to the Earth first. Here we check this supposition. For this purpose, along with AGASA data[1] we have considered Yakutsk EAS array data beginning from $E>5.10^{18}$ eV. Data for the period of 1974-2002 are used and the arrival directions of $< 60°$ zenith angle EAS (n=1863) have been analysis. Among them there are 5157 doublets (clusters having the great number of particles, triplet and etc, are not considered). The distance between the EAS in doublets is $< 5°$.

The total number of doublets is separated into 2 parts: doublets $N_1(E_1)$ whose particle having the energy $E_1$ greater than the energy $E_2$ of neighboring particle ($E_1 > E_2$, doublet has 2 particles with energies $E_1, E_2$) arrives first to the Earth, $N_2(E_1)$ are remaining doublets. It is known that the shower energy is determined with a some error ~ 30%. Therefore, doublets, whose particles have energies close to each other within $\log(E_1/E_2) < 0.2$, are excluded.

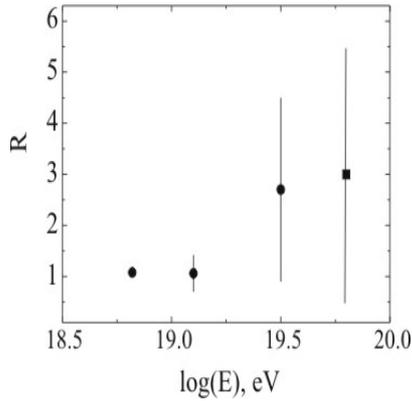 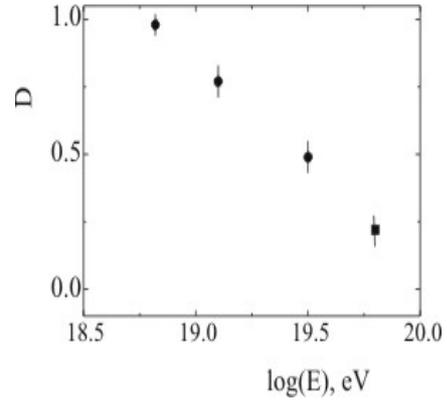

*Figure 1.* Ratio of the number of doublets $R=N_1(E_1)/N_2(E_1)$

*Figure 2.* A portion of EAS forming doublets $N_1$ to the total number of EAS $N_2$.: $D=N_1/N_2$

A ratio of the number of these doublets is presented in Fig.1 (circles –Yakutsk, squares – AGASA): $R=N_1(E_1)/N_2(E_1)$, where $N_1(E_1)$ are doublets whose particle of $E_1$ ($E_1>E_2$) arrives to the Earth first, $N_2(E_1)$ - rest doublets.

The doublet[1] A6 has been excluded from AGASA data because the doublet particles have close to each other energies. As seen from Fig.1, at $E<2.10^{19}$ eV doublets arrive uniformly on the average independent of the energy of leading particles. At $E>2.10^{19}$ eV the number of doublets, whose particle of $E_1$ is a leader, is ~70% for Yakutsk array and ~80% for AGASA. So, one can conclude that doublets of $E>2.10^{19}$ eV are most likely formed as a result of super heavy nuclei fragmentation. Apparently, at $E<2.10^{19}$ eV doublets are formed by other means than doublets of the higher energies.

Fig.2 presents a portion of EAS forming doublets relative to the total number of EAS: $D = N_1/N_2$, where $N_1$ is the number of EAS forming doublets, $N_2$ is the total number of EAS. As seen from Fig.2, a portion of EAS forming doublets decreases with energy by Yakutsk data.

Further, we considered the distribution of doublets in galactic latitude through 3° for three energy intervals (Fig.3): a) $10^{18.7}$–$10^{19}$ eV – n=4822 doublets (1407 EAS), b) $10^{19}$ – $10^{19.3}$ eV – n=293 (337 EAS), c) $>10^{19.3}$ eV– n=42 (59 EAS). The histogram presents averaged latitudes for particles of doublets. In Fig.3a for $E=10^{18.7}$–$10^{19}$ eV the galactic plane is seen. At latitudes $|b|<3°$ the excess of the number of doublets observed relative to the number expected ones in the case of isotropy is $11\sigma=(474-288.9)/\sqrt{288.9}$. The number of doublets expected is found by the Monte-Carlo method taking into account the exposure of array on the celestial sphere. At latitudes of $24°>b>9°$ a maximum in the doublet distribution is also observed. More detail consideration shows that this maximum is limited by $150°>b>120°$ in galactic longitude. This maximum in doublet distribution repeats a maximum in the distribution of EAS for same coordinates at $E\sim10^{19}$ eV (Fig.2 of paper[3]).

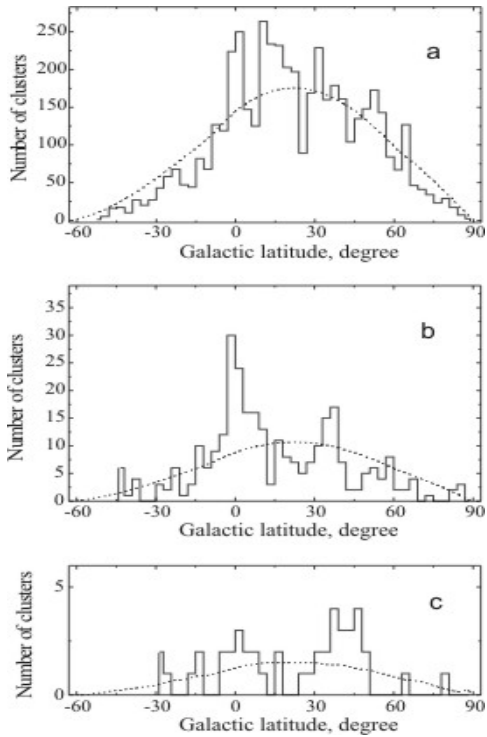

*Figure 3.* The distribution of doublets in latitude the energy interval: a) $10^{18.7}$–$10^{19}$ eV, b) $10^{19}$ – $10^{19.3}$ eV, c) E>$10^{19.3}$ eV.

At │b│<3° for $10^{19}$-$10^{19.3}$ eV the number of doublets exceeds the number of doublets expected by 8.7σ=(54-17.5)/$\sqrt{17.5}$. Note that early we found[4] the flux of particles from the galactic plane side at E=$10^{18.9}$-$10^{19.6}$ eV exceeds by 4.2σ the expected ones. The distribution of doublets at E>$10^{19.3}$ eV (Fig.3c) is more isotropic. So, the arrival direction of doublets (clusters) can be an indicator of cosmic rays anisotropy.

Thus, the distribution of doublets in galactic latitude indicates that at E~$10^{19}$ eV cosmic rays are galactic. This result confirm conclusions obtained by us earlier using other methods[4,5]. As indicated above, at this energy ~ 80% of showers form doublets. From this it follows that the doublets are mainly the ordinary EAS they are not formed by neutral particles.

In the galactic model for the cosmic ray origin, in the case of protons from the galactic plane side a broad maximum of ~ 40° is expected and for the iron nuclei it is the uniform distribution of particle fluxes in latitude[6]. The narrow maximum in the distribution of doublets from the galactic plane side at ~ $10^{19}$ eV (Fig.3b) can testify that the primaries are iron nuclei (probably, the regular magnetic field absent in the galactic plane and the iron nuclei can propagate along it) as it concluded in[6].

From the above analysis we conclude:
1) doublets (clusters) at E<$2.10^{19}$ eV are on whole the ordinary EAS,
2) clusters at E>$2.10^{19}$ eV are most likely formed by means of super heavy nuclei fragmentation,
3) a portion of EAS forming clusters decreases with energy,
4) the distribution of doublets in galactic latitude indicators that cosmic rays at E~$10^{19}$ eV consist most likely of iron nuclei and are galactic,
5) the arrival directions of doublets can be indicators of the cosmic ray anisotropy.

**ACKNOWLEDGEMENTS** This work is supported by Russian FBR (grant 04-02-16287).